\documentclass[11pt]{article}
\usepackage{graphicx}
\usepackage{amsmath}
\usepackage{harvard}
\usepackage{a4}

\input{tcilatex}

\begin{document}

\title{Slowly rotating, compact fluid sources embedded in Kerr empty
space-time}
\author{Ron Wiltshire, \\
Division of Mathematics and Statistics,\\
The University of Glamorgan,\\
Pontypridd, CF37 1DL.\\
email: rjwiltsh@glam.ac.uk}
\maketitle

\begin{abstract}
Spherically symmetric static fluid sources are endowed with rotation and
embedded in Kerr empty space-time up to an including quadratic terms in an
angular velocity parameter using Darmois junction conditions. Einstein's
equation's for the system are developed in terms of linear ordinary
differential equations. The boundary of the rotating source is expressed
explicitly in terms of sinusoidal functions of the polar angle which differ
somewhat according to whether an equation of state exists between internal
density and supporting pressure.
\end{abstract}

Following the publication by Kerr [1] of the metric which describes
analytically, the asymptotically flat, vacuum gravitational field outside a
rotating source in terms of Einstein's field equations, there has been much
discussion concerning the existence of possible interior solutions which
match the exterior smoothly. In an important development Hartle [2] uses a
second order perturbation technique to describe the slow rotation of
equilibrium configurations of cold stars having constant angular velocity.
Solutions of Einstein's equations are developed in terms of Legendre
polynomials but the issue of matching the results to Kerr empty space-time
is not addressed. In the case of non-equilibrium configurations Kegeles [3]
has applied the method to Robertson-Walker dust sources up to the first
order in angular velocity parameter although, the results are somewhat
restrictive and are not suitable for application to sources supported by
internal pressure. In a recent work the case of the Wahlquist [4] closed
form interior was shown not to fit the Kerr exterior by Bradley\textit{\ et
al }[5]. Only for the important case \ of thin super-massive rotating discs,
supported by internal pressure have analytic sources for the Kerr metric
been found (Pichon and Lynden-Bell [6]). This has led to an `embarrassing \
hiatus' according to Bradley \textit{et al} [5] in the number of \ potential
interior solutions available for matching which \ in turn has contributed to
a lack in the development in the theory of differentially rotating fluid
bodies in general relativity. Yet it is important to develop further the
relativistic theory of rotation since it has considerable potential
application in astrophysics, for example, in the description of the
gravitational collapse of rotating matter, quasars, or potential sources for
gravitational radiation.

It is the aim here to focus upon one aspect of the problem of matching
rotating interior solutions of Einstein's equations to the Kerr metric. In
particular it is the intention to show how static perfect fluid sources
endowed with rotation to an accuracy of second order terms in an angular
velocity parameter $q$ may be matched to the Kerr metric also expressed to
an accuracy of quadratic terms in angular velocity parameter. In this sense
the method employed is similar to that of Bradley \textit{et al }[5].
Moreover, the results given here apply to fluid interiors having variable
interior angular velocity and so to that extent generalise the discussion of
Hartle [2]. Indeed the problem considered here is essentially that of
matching the generalised Hartle rotating interiors to Kerr empty space-time
although the approach here differs significantly. The matching problem is
addressed by means of the application of Darmois [7] junction conditions
which are discussed widely in the literature for example, Misner et al [8],
Mars \& Senovilla [9], Stephani [10], Hernandez-Pastora et al [11] and
Bonnor and Vickers [12]. In the Darmois approach it is necessary that the
components of the metric tensor, and also the extrinsic curvature for the
Kerr exterior and the interior source are continuous at the boundary surface.

In the following Einstein's equations for a perfect fluid source will be
expressed in the form:

\begin{equation}
G_{b}^{a}=-8\pi T_{b}^{a}\quad \text{,}\quad \quad T_{b}^{a}=\left( \rho
+p\right) u^{a}u_{b}-\delta _{b}^{a}p\quad \text{,}  \label{ebo60p}
\end{equation}%
where $\rho $, $p$ are the respective rotating source density and supporting
internal pressure and $u^{a}$ are the components of the velocity four-vector
with the property that $u^{a}u_{a}=1.$

\bigskip

\section{The fluid source and Darmois junction conditions with empty
space-time}

Consider the well known spherically static perfect fluid source which is
represented here by the metric:%
\begin{equation}
d\sigma _{S}^{2}=e^{2\lambda }d\eta ^{2}-e^{2\mu }d\xi ^{2}-\xi ^{2}d\theta
^{2}-\xi ^{2}\sin ^{2}\theta d\phi ^{2}  \label{stat5}
\end{equation}%
where $\lambda =\lambda \left( \xi \right) $ and $\mu =\mu \left( \xi
\right) $ such that 
\begin{equation}
\lambda _{\xi \xi }=-\lambda _{\xi }^{2}+\mu _{\xi }\lambda _{\xi }+\frac{%
\lambda _{\xi }+\mu _{\xi }}{\xi }-\frac{e^{2\mu }}{\xi ^{2}}+\frac{1}{\xi
^{2}}  \label{stat10}
\end{equation}%
and a suffix indicates a derivative. The respective supporting pressure and
density of the static cases are:%
\begin{equation}
8\pi p_{S}=\frac{2\,e^{-{2\,\mu }}\,\lambda _{\xi }}{\xi }+\frac{e^{-{2\,\mu 
}}}{\xi ^{2}}-\frac{1}{\xi ^{2}}  \label{stat15}
\end{equation}%
and%
\begin{equation}
8\pi \rho _{S}=\frac{2\,e^{-{2\,\mu }}\,\mu _{\xi }}{\xi }-\frac{e^{-{2\,\mu 
}}}{\xi ^{2}}+\frac{1}{\xi ^{2}}  \label{stat20}
\end{equation}

In the following this source will be endowed with rotation up to and
including quadratic terms in angular velocity parameter $q$. It is assumed
that the resulting physical system may be described by means of the metric%
\begin{eqnarray}
d\sigma ^{2} &=&e^{2\lambda }\left( 1+Qq^{2}\right) d\eta ^{2}-e^{2\mu
}\left( 1+Uq^{2}\right) d\xi ^{2}-\xi ^{2}\left( 1+Vq^{2}\right) d\theta ^{2}
\notag \\
&&-\xi ^{2}\left( 1+Wq^{2}\right) \sin ^{2}\theta d\phi ^{2}-2X\xi ^{2}q\sin
^{2}\theta d\phi d\eta  \label{tech5}
\end{eqnarray}%
where $U$,$V$, $W$ and $Q$ are each functions of $\xi $ and $\theta $ whilst 
$X$ is a function of $\xi $ alone. This particular gauge has been chosen
since it is the intention to match this \ interior source to empty
space-time using the well known Kerr metric expressed in Boyer and Lindquist
coordinates \ at a boundary between source and empty space-time given by:%
\begin{equation}
\xi _{b}=\xi _{0}+q^{2}f\left( \theta \right)  \label{tech10}
\end{equation}%
where $f=f\left( \theta \right) $ and $\xi _{0}$ is constant. All
expressions will be accurate up to and including quadratic terms in $q$ so
that the Kerr metric may be written as:%
\begin{equation}
d\sigma ^{2}=\Psi _{44}d\eta ^{2}-2\Psi _{43}q\sin ^{2}\theta d\phi d\eta
-\Psi _{11}d\xi ^{2}-\Psi _{22}d\theta ^{2}-\Psi _{33}\sin ^{2}\theta d\phi
^{2}  \label{ebo300}
\end{equation}%
where:%
\begin{eqnarray}
\Psi _{44} &=&1-\frac{2m}{\xi }\left( 1-\frac{q^{2}\cos ^{2}\theta }{\xi ^{2}%
}\right)  \notag \\
\Psi _{43} &=&\frac{2m}{\xi }  \notag \\
\Psi _{11} &=&\frac{1}{1-\frac{2m}{\xi }}\left[ 1+\frac{q^{2}\cos ^{2}\theta 
}{\xi ^{2}}-\frac{q^{2}}{\xi ^{2}\left( 1-\frac{2m}{\xi }\right) }\right] 
\notag \\
\Psi _{22} &=&\xi ^{2}+q^{2}\cos ^{2}\theta  \notag \\
\Psi _{33} &=&\frac{2mq^{2}\sin ^{2}\theta }{\xi }+q^{2}+\xi ^{2}
\label{ebo302}
\end{eqnarray}

Darmois junction conditions are employed at the boundary so that both the
metric tensor $g_{ab}$ and the extrinsic curvature $K_{ab}$ must be
continuous on the bondary surface (\ref{tech10}). Using the suffix $b$ to
indicate evaluation at $\xi =\xi _{b}$ and the suffix $0$ to indicate
evaluation at $\xi =\xi _{0}$ it is straight forward to show that the
following conditions on $\lambda $, $\mu $, $U$, $V$, $W$ , $Q$ and $X$ and
derivatives with respect to $\xi $ hold at the boundary:

\begin{eqnarray}
\left\{ e^{2\lambda }\right\} _{0} &=&\left\{ e^{-2\mu }\right\} _{0}=1-%
\frac{2m}{\xi _{0}}\equiv K  \notag \\
\left\{ \lambda _{\xi }\right\} _{0} &=&\frac{1-K}{2K\xi _{0}}  \notag \\
U_{b} &=&\frac{K\cos ^{2}\theta -1}{K\xi _{0}^{2}}+2f\left\{ \mu _{\xi
}\right\} _{0}+\frac{f\left( K-1\right) }{K\xi _{0}}  \notag \\
V_{b} &=&\frac{\cos ^{2}\theta }{\xi _{0}^{2}}  \notag \\
\left\{ V_{\xi }\right\} _{b} &=&\frac{4f\left\{ \mu _{\xi }\right\} _{0}}{%
\xi _{0}}-\frac{2\cos ^{2}\theta }{\xi _{0}^{3}}  \notag \\
W_{b} &=&\frac{1+\left( 1-K\right) \sin ^{2}\theta }{\xi _{0}^{2}}  \notag \\
\left\{ W_{\xi }\right\} _{b} &=&\frac{3\left( K-1\right) \sin ^{2}\theta -2%
}{\xi _{0}^{3}}+\frac{4f\left\{ \mu _{\xi }\right\} _{0}}{\xi _{0}}  \notag
\\
Q_{b} &=&\frac{\left( 1-K\right) \cos ^{2}\theta }{K\xi _{0}^{2}}  \notag \\
\left\{ Q_{\xi }\right\} _{b} &=&\frac{\left( 2K^{2}-K-1\right) \cos
^{2}\theta }{\xi _{0}^{3}K^{2}}+\frac{f\left( 1-3K\right) \left\{ \mu _{\xi
}\right\} _{0}}{\xi _{0}K}+\frac{f\left( K^{2}-1\right) }{2\xi _{0}^{2}K^{2}}
\notag \\
X_{b} &=&\frac{1-K}{\xi _{0}^{2}}  \notag \\
\left\{ X_{\xi }\right\} _{b} &=&\frac{3\left( K-1\right) }{\xi _{0}^{3}}
\label{conds5}
\end{eqnarray}%
For future convience it is useful to define the functions $H\left( \xi
,\theta \right) $ and $h\left( \xi \right) $ such that:%
\begin{eqnarray}
H\left( \xi ,\theta \right)  &=&Q\left( \xi ,\theta \right) +\sin ^{2}\theta
\xi ^{2}X^{2}e^{-2\lambda }  \notag \\
X_{\xi } &=&\frac{h\left( \xi \right) e^{\lambda +\mu }}{\xi ^{4}}
\label{red8}
\end{eqnarray}%
so:%
\begin{eqnarray}
H_{b} &=&\frac{\left( 1-K\right) }{K\xi _{0}^{2}}\left( 1-K\sin ^{2}\theta
\right)   \notag \\
\left\{ H_{\xi }\right\} _{b} &=&\frac{3\sin ^{2}\theta \left( 1-K\right) }{%
\xi _{0}^{3}}+\frac{\left( 2K^{2}-K-1\right) }{K^{2}\xi _{0}^{3}}+\frac{%
f\left( K^{2}-1\right) }{2\xi _{0}^{2}K^{2}}  \notag \\
&&+\frac{f\left\{ \mu _{\xi }\right\} _{0}\left( 1-3K\right) }{\xi _{0}K} 
\notag \\
h_{b} &=&3\xi _{0}\left( K-1\right)   \label{red17}
\end{eqnarray}

\newpage

\section{Einstein's equations for the \ rotating source}

There are three non-trivial Einstein equations (\ref{ebo60p}) which need to
be solved for a fluid source in the gauge (\ref{tech5}). In the first the
condition $T_{2}^{1}=0=T_{1}^{2}$ gives rise to:%
\begin{equation}
-U_{\theta }\lambda _{\xi }\,+\,H_{\theta }\lambda _{\xi }+W_{\theta \xi }-%
\frac{U_{\theta }}{\xi }+H_{\theta \xi }-\frac{H_{\theta }}{\xi }+\frac{\cos
\theta \,\left( W_{\xi }-V_{\xi }\right) }{\sin \theta }=0  \label{e12}
\end{equation}%
whilst $T_{1}^{1}=T_{2}^{2}$ becomes:%
\begin{eqnarray}
&&\frac{e^{-{2\,\mu }}}{2}\left\{ -V_{\xi }\lambda _{\xi }-U_{\xi }\lambda
_{\xi }+2H_{\xi }\lambda _{\xi }+W_{\xi \xi }+\frac{\,\,W_{\xi }\,}{\xi }%
-\mu _{\xi }W_{\xi }\right.  \notag \\
&&\left. -\frac{V_{\xi }}{\xi }-\frac{\,U_{\xi }}{\,\xi }+H_{\xi \xi }-\frac{%
H_{\xi }}{\,\xi }-\mu _{\xi }H_{\xi }\right\}  \notag \\
&&+\frac{\,\cos \theta \,\left( U_{\theta }+V_{\theta }-2W_{\theta }\right) 
}{2\xi ^{2}\sin \theta }-\frac{H_{\theta \theta }+W_{\theta \theta }}{2\xi
^{2}}+\frac{U-V}{\xi ^{2}}  \notag \\
&&-\frac{h^{2}\,\sin ^{2}\theta }{2\,\xi ^{6}}  \notag \\
&=&0  \label{e1122}
\end{eqnarray}%
The final equation is $\left( T_{3}^{3}+p\right) \left( T_{4}^{4}+p\right)
-T_{4}^{3}T_{3}^{4}=0$ which is explicitly:%
\begin{eqnarray}
&&e^{-{2\,\mu }}\,\left( \lambda _{\xi }+\mu _{\xi }\right) \left\{ -W_{\xi
\xi }-\lambda _{\xi }\,W_{\xi }-\frac{2\,W_{\xi }}{\xi }+\mu _{\xi }\,W_{\xi
}\,+V_{\xi \xi }\right.  \notag \\
&&\left. +\lambda _{\xi }\,V_{\xi }+\frac{2V_{\xi }}{\xi }-\mu _{\xi
}\,V_{\xi }\,\right\} -\frac{\cos \theta \,\left( H_{\theta }+U_{\theta
}\right) \left( \lambda _{\xi }+\mu _{\xi }\right) }{\xi ^{2}\sin \theta } 
\notag \\
&&+\frac{\left( H_{\theta \theta }+U_{\theta \theta }\right) \left( \lambda
_{\xi }+\mu _{\xi }\right) }{\xi ^{2}}-\sin ^{2}\theta \,\left( \frac{h_{\xi
}^{2}}{4\,\xi ^{5}\,}+\frac{\left( \lambda _{\xi }+\mu _{\xi }\right) h^{2}}{%
\xi ^{6}}\right)  \notag \\
&=&0  \label{e43}
\end{eqnarray}

An important differential consequence of these equations, since it is
independent of any second derivatives with respect to $\xi $ is the relation:%
\begin{eqnarray}
&&\frac{e^{-{2\,\mu }}\sin \theta }{\cos \theta }\left( U_{\theta
}-H_{\theta }\right) \xi ^{2}\lambda _{\xi }\left( \lambda _{\xi }\,-\mu
_{\xi }\right) \left( \lambda _{\xi }+\mu _{\xi }\right)   \notag \\
&&+e^{-{2\,\mu }}\left( V_{\xi }\,-W_{\xi }\,\right) \left( \,\xi
^{2}\lambda _{\xi }-\,\mu _{\xi }\xi ^{2}+2\xi \right) \left( \lambda _{\xi
}+\mu _{\xi }\right)   \notag \\
&&+\frac{e^{-{2\,\mu }}\,\lambda _{\xi }\sin \theta \,\,}{\cos \theta }%
\left( \xi ^{2}\,V_{\theta \xi }-\xi \,U_{\theta }-\xi ^{2}\,H_{\theta \xi
}+\xi \,H_{\theta }\right) \left( \lambda _{\xi }+\mu _{\xi }\right)   \notag
\\
&&+\frac{\sin \theta \,}{\cos \theta }\left( W_{\theta \theta \theta
}-2\,W_{\theta }+3\,V_{\theta }+H_{\theta \theta \theta }-H_{\theta }\right)
\left( \lambda _{\xi }+\mu _{\xi }\right)   \notag \\
&&+\frac{e^{-{2\,\mu }}\,\left( \lambda _{\xi }+\mu _{\xi }\right) \sin
\theta }{\cos \theta }\left\{ \mu _{\xi }\,\left( \xi ^{2}W_{\theta \xi
}\,-\xi \,U_{\theta }+\xi ^{2}\,H_{\theta \xi }+\xi \,H_{\theta }\right)
\right.   \notag \\
&&\left. -\xi \,W_{\theta \xi }+\xi V_{\theta \xi }\,+2H_{\theta }\,\right\}
+\frac{\cos \theta \,}{\sin \theta }\left( -2\,W_{\theta }+V_{\theta
}-H_{\theta }\right) \left( \lambda _{\xi }+\mu _{\xi }\right)   \notag \\
&&+\left( 2W_{\theta \theta }-V_{\theta \theta }+H_{\theta \theta }\right)
\left( \lambda _{\xi }+\mu _{\xi }\right) +\sin ^{2}\theta \left( \frac{%
\left( \lambda _{\xi }+\mu _{\xi }\right) h^{2}\,}{\xi ^{4}}-\frac{h_{\xi
}^{2}\,}{4\,\xi ^{3}\,}\right)   \notag \\
&=&0  \label{conse}
\end{eqnarray}%
The expression for the supporting internal pressure is:%
\begin{eqnarray}
8\pi p &=&8\pi p_{S}+e^{-{2\,\mu }}\,q^{2}\,\left( \frac{W_{\xi }\lambda
_{\xi }}{2}+\frac{V_{\xi }\lambda _{\xi }\,}{2}-\frac{2\,U\,\lambda _{\xi }}{%
\xi }+\frac{W_{\xi }}{2\,\xi }+\frac{V_{\xi }}{2\,\xi }-\frac{U}{\xi ^{2}}+%
\frac{H_{\xi }}{\xi }\right)   \notag \\
&&+\frac{q^{2}}{2\xi ^{2}}\,\left( W_{\theta \theta }+2V+H_{\theta \theta
}\right)   \notag \\
&&+\frac{q^{2}\,\cos \theta \,\left( 2W_{\theta }-V_{\theta }+H_{\theta
}\right) }{2\xi ^{2}\sin \theta }+\frac{h^{2}\,q^{2}\,\sin ^{2}\theta }{%
4\,\xi ^{6}}  \label{pres}
\end{eqnarray}%
and the fluid density is:

\begin{eqnarray}
&&8\pi \rho =8\pi \rho _{S}+e^{-{2\,\mu }}\,q^{2}\left\{ \frac{W_{\xi
}\lambda _{\xi }}{2}-\frac{V_{\xi }\,\lambda _{\xi }}{2}-\frac{W_{\xi }}{%
2\,\xi }-V_{\xi \xi }-\frac{5V_{\xi }}{2\,\xi }+\mu _{\xi }V_{\xi }\right. 
\notag \\
&&\left. +\frac{U_{\xi }}{\xi }-\frac{2\,\mu _{\xi }\,U}{\xi }+\frac{U}{\xi
^{2}}\right\} -\frac{q^{2}}{2\xi ^{2}}\,\left( W_{\theta \theta
}+2V+2U_{\theta \theta }+H_{\theta \theta }\right)  \notag \\
&&+\frac{q^{2}\,\cos \theta \,\left( -2W_{\theta }+V_{\theta }+H_{\theta
}\right) }{2\xi ^{2}\sin \theta }+\frac{h^{2}\,q^{2}\,\sin ^{2}\theta }{%
4\,\xi ^{6}}  \label{dens}
\end{eqnarray}%
The angular velocity may be written as:%
\begin{equation}
L\left( \xi ,\eta \right) \equiv \frac{u^{3}}{u^{4}}=\frac{T_{4}^{3}}{%
T_{4}^{4}+p}=\frac{qh_{\xi }e^{\lambda +\mu }}{4\left( \lambda _{\xi }+\mu
_{\xi }\right) \xi ^{3}}-qX  \label{ang}
\end{equation}

\section{Description of the boundary}

Application of the boundary conditions (\ref{conds5}) and (\ref{red17}) show
that equation (\ref{e12}) is satisfied identically whilst (\ref{pres}) gives:%
\begin{equation}
8\pi p_{b}=0  \label{boun5}
\end{equation}%
as expected.

Whenever $\lambda _{\xi }+\mu _{\xi }\neq 0$, so that from the second of (%
\ref{conds5}) 

\begin{equation}
J=2\xi _{0}\left\{ \mu _{\xi }\right\} _{0}K-K+1\neq 0  \label{bou5}
\end{equation}%
then equation (\ref{conse}) may be used  to define $f\left( \theta \right) $
since at the boundary it reduces to:%
\begin{eqnarray}
&&2\cos \theta \sin \theta KJ\left( 4\xi _{0}\left\{ \mu _{\xi }\right\}
_{0}K\left( K-1\right) +7K^{2}-10K-1\right)   \notag \\
&&-\frac{\xi _{0}f_{\theta }J}{2K}\left( 2\xi _{0}\left\{ \mu _{\xi
}\right\} _{0}K+K-1\right) \left( 2\xi _{0}\left\{ \mu _{\xi }\right\}
_{0}K\left( K+1\right) -3K^{2}+4K-1\right)   \notag \\
&&-\cos \theta \sin \theta K^{2}\left\{ h_{\xi }^{2}\right\} _{b}  \notag \\
&=&0  \label{boun16}
\end{eqnarray}%
Thus in such cases it follows that:%
\begin{equation}
f\left( \theta \right) =a_{0}+a_{1}\sin ^{2}\theta   \label{boun15}
\end{equation}%
where $a_{0}$ is an arbitrary constant and $a_{1}$ is found using (\ref%
{boun16}).

However the case $\left\{ \lambda _{\xi }+\mu _{\xi }\right\} _{0}=0$ is
also of importance. From equation (\ref{conse}) it follows that  $\left\{
h_{\xi }\right\} _{b}=0$ and so the value of the angular velocity $L\left(
\xi _{b},\eta \right) $ at the boundary surface must be determined by
careful analysis of the limiting behaviour of (\ref{ang}) at $\xi =\xi _{b}$%
. Moreover since this case corresponds to the condition $J=0$ then the
second of (\ref{red17}) and (\ref{stat20}) imply that the surface density
for the static solution must satisfy $\left\{ 8\pi \rho _{S}\right\} _{0}=0$%
. This will invariably be the case in physical situations where an equation
of state of the type $p=F\left( \rho \right) $ is additionally imposed on
equations (\ref{e12}) to (\ref{e43}) to close the system of Einstein's
equations. For such cases the condition $\left\{ 8\pi \rho \right\} _{b}=0$
also gives:

\begin{eqnarray}
&&f\left( \frac{2K^{2}\left\{ \mu _{\xi \xi }\right\} _{0}}{\xi _{0}}-\frac{%
10K^{2}\left\{ \mu _{\xi }\right\} _{0}}{\xi _{0}^{2}}+\frac{K^{2}+K-2}{\xi
_{0}^{3}}\right)  \notag \\
&&+K^{2}\left( -\left\{ V_{\xi \xi }\right\} _{0}+\frac{\left\{ U_{\xi
}\right\} _{0}}{\xi _{0}}+\frac{8\cos ^{2}\theta }{\xi _{0}^{4}}\right) 
\notag \\
&&+f_{\theta \theta }\left( -\frac{2K\left\{ \mu _{\xi }\right\} _{0}}{\xi
_{0}^{2}}+\frac{1-K}{\xi _{0}^{3}}\right) -\frac{1+K}{\xi _{0}^{4}}  \notag
\\
&=&0  \label{boun21}
\end{eqnarray}

This equation may be used to define the equation of the boundary $f=f\left(
\theta \right) $ whenever $\ \left\{ U_{\xi }\right\} _{0}$ and $\left\{
V_{\xi \xi }\right\} _{0}$ are known. Clearly whenever $\left\{ U_{\xi
}\right\} _{0}$ and $\left\{ V_{\xi \xi }\right\} _{0}$ depend purely on
terms in $\sin ^{2}\theta $ then $f\left( \theta \right) $ will have a
solution of the form (\ref{boun15}) and $a_{0}$ and $a_{1}$ would be found
using (\ref{boun21}).

\section{Generating solutions of Einstein's equations }

Consider now the condition $\left\{ \lambda _{\xi }+\mu _{\xi }\right\}
_{0}\neq 0$ and also the condition $\left\{ \lambda _{\xi }+\mu _{\xi
}\right\} _{0}=0$ provided that $\left\{ U_{\xi }\right\} _{0}$ and $\left\{
V_{\xi \xi }\right\} _{0}$ depend purely on terms in $\sin ^{2}\theta $. The
case when $\left\{ U_{\xi }\right\} _{0}$ and $\left\{ V_{\xi \xi }\right\}
_{0}$ depend on higher harmonics is discussed in the next section.

It follows that $f\left( \theta \right) $ has the form (\ref{boun15}) and
for such cases solutions of Einstein's equations satisfying the boundary
Darmois boundary conditions may be found by writing:%
\begin{eqnarray}
U\left( \xi ,\theta \right)  &=&G_{0}\left( \xi \right) +G_{1}\left( \xi
\right) \sin ^{2}\theta   \notag \\
V\left( \xi ,\theta \right)  &=&G_{2}\left( \xi \right) +G_{3}\left( \xi
\right) \sin ^{2}\theta   \notag \\
W\left( \xi ,\theta \right)  &=&G_{2}\left( \xi \right) +G_{5}\left( \xi
\right) \sin ^{2}\theta   \notag \\
H\left( \xi ,\theta \right)  &=&G_{6}\left( \xi \right) +G_{7}\left( \xi
\right) \sin ^{2}\theta   \label{gees}
\end{eqnarray}

The equations (\ref{e12}), (\ref{e1122}) and (\ref{e43}) then give rise to
four independent equations for $G_{i}\left( \xi \right) $ as follows:%
\begin{equation}
2\,G_{7}\,\lambda _{\xi }-2\,G_{1}\,\lambda _{\xi }-\frac{2\,G_{7}}{\xi }-%
\frac{2\,G_{1}}{\xi }+2\,G_{7_{\xi }}+3\,G_{5_{\xi }}-G_{3_{\xi }}=0
\label{eqn5}
\end{equation}%
\begin{eqnarray}
&&2\,G_{7_{\xi }}\,\lambda _{\xi }-\,G_{3_{\xi }}\lambda _{\xi }-G_{1_{\xi
}}\lambda _{\xi }-\frac{G_{7_{\xi }}}{\xi }+\frac{G_{5_{\xi }}}{\xi }-\frac{%
G_{3_{\xi }}}{\xi }-\frac{G_{1_{\xi }}}{\xi }  \notag \\
&&+\frac{4\,G_{7}\,e^{2\,\mu }}{\xi ^{2}}+\frac{8\,G_{5}\,e^{2\,\mu }}{\xi
^{2}}-\frac{4\,G_{3}\,e^{2\,\mu }}{\xi ^{2}}-\frac{h^{2}\,e^{2\,\mu }}{\xi
^{6}}  \notag \\
&&-G_{7_{\xi }}\mu _{\xi }-\,G_{5_{\xi }}\mu _{\xi }+G_{7_{\xi \xi
}}+G_{5_{\xi \xi }}  \notag \\
&=&0  \label{eqn7}
\end{eqnarray}%
\begin{eqnarray}
&&2G_{7}\,\left( \,\lambda _{\xi }^{2}-\frac{\lambda _{\xi }\,}{\xi }-\mu
_{\xi }\,\lambda _{\xi }-\frac{\mu _{\xi }}{\xi }+\frac{6\,e^{2\,\mu }}{\xi
^{2}}-\frac{2}{\xi ^{2}}\right) +  \notag \\
&&2G_{1}\,\left( -\,\lambda _{\xi }^{2}+\frac{\lambda _{\xi }}{\xi }+\mu
_{\xi }\lambda _{\xi }+\frac{\mu _{\xi }}{\xi }\right) +\frac{h_{\xi
}^{2}\,e^{2\,\mu }}{4\,\xi ^{5}\left( \lambda _{\xi }+\mu _{\xi }\right) } 
\notag \\
&&+2G_{7_{\xi }}\,\left( \lambda _{\xi }-\,\mu _{\xi }\right) +\,G_{5_{\xi
}}\left( \lambda _{\xi }+\frac{4}{\xi }-3\mu _{\xi }\right)  \notag \\
&&+G_{3_{\xi }}\,\left( -3\lambda _{\xi }-\frac{4}{\xi }+\mu _{\xi }\right) +%
\frac{16\,G_{5}\,e^{2\,\mu }}{\xi ^{2}}-\frac{8\,G_{3}\,e^{2\,\mu }}{\xi ^{2}%
}-\frac{h^{2}\,e^{2\,\mu }}{\xi ^{6}}  \notag \\
&=&0  \label{eqn10}
\end{eqnarray}%
\begin{eqnarray}
&&-2\,G_{6_{\xi }}\lambda _{\xi }+G_{2_{\xi }}\lambda _{\xi }+G_{0_{\xi
}}\lambda _{\xi }+\frac{G_{6_{\xi }}}{\xi }+\frac{G_{0_{\xi }}}{\xi }  \notag
\\
&&+\frac{2\left( \,G_{7}+3\,G_{5}-\,G_{3}+\,G_{2}-\,G_{1}-\,G_{0}\right)
\,e^{2\,\mu }}{\xi ^{2}}  \notag \\
&&+\,G_{6_{\xi }}\mu _{\xi }+\,G_{2_{\xi }}\mu _{\xi }-G_{6_{\xi \xi
}}-G_{2_{\xi \xi }}  \notag \\
&=&0  \label{eqn25}
\end{eqnarray}

The first three contain only $G_{1}\left( \xi \right) $, $G_{3}\left( \xi
\right) $, $G_{5}\left( \xi \right) $ and $G_{7}\left( \xi \right) $ whilst
the fourth contains each of the unknown $G_{i}\left( \xi \right) $. This
under-determined set may be closed with the aid of an imposed equation of
state as discussed in the previous section.

The boundary conditions (\ref{conds5}) and (\ref{red17}) together with (\ref%
{gees}) then give rise to the following:%
\begin{eqnarray}
G_{0}\left( \xi _{0}\right) &=&\frac{K-1}{K}\left( \frac{1}{\xi _{0}^{2}}+%
\frac{a_{0}}{\xi _{0}}\right) +2a_{0}\left\{ \mu _{\xi }\right\} _{0}  \notag
\\
G_{1}\left( \xi _{0}\right) &=&2a_{1}\left\{ \mu _{\xi }\right\} _{0}-\frac{1%
}{\xi _{0}^{2}}-\frac{a_{1}\left( 1-K\right) }{\xi _{0}K}  \notag \\
G_{2}\left( \xi _{0}\right) &=&\frac{1}{\xi _{0}^{2}}  \notag \\
G_{3}\left( \xi _{0}\right) &=&-\frac{1}{\xi _{0}^{2}}  \notag \\
G_{5}\left( \xi _{0}\right) &=&\frac{1-K}{\xi _{0}^{2}}  \notag \\
G_{6}\left( \xi _{0}\right) &=&\frac{1-K}{K\xi _{0}^{2}}  \notag \\
G_{7}\left( \xi _{0}\right) &=&\frac{K-1}{\xi _{0}^{2}}  \label{ser16}
\end{eqnarray}%
and:%
\begin{eqnarray}
\qquad G_{2_{\xi }}\left( \xi _{0}\right) &=&\frac{4\left\{ \mu _{\xi
}\right\} _{0}a_{0}}{\xi _{0}}-\frac{2}{\xi _{0}^{3}}  \notag \\
G_{3_{\xi }}\left( \xi _{0}\right) &=&\frac{4\left\{ \mu _{\xi }\right\}
_{0}a_{1}}{\xi _{0}}+\frac{2}{\xi _{0}^{3}}  \notag \\
G_{5_{\xi }}\left( \xi _{0}\right) &=&\frac{4\left\{ \mu _{\xi }\right\}
_{0}a_{1}}{\xi _{0}}+\frac{3\left( K-1\right) }{\xi _{0}^{3}}  \notag \\
G_{6_{\xi }}\left( \xi _{0}\right) &=&\frac{a_{0}\left\{ \mu _{\xi }\right\}
_{0}}{\xi _{0}K}-\frac{1}{\xi _{0}^{3}K}-\frac{a_{0}}{2\xi _{0}^{2}K^{2}}-%
\frac{1}{\xi _{0}^{3}K^{2}}  \notag \\
&&-\frac{3a_{0}\left\{ \mu _{\xi }\right\} _{0}}{\xi _{0}}+\frac{a_{0}}{2\xi
_{0}^{2}}+\frac{2}{\xi _{0}^{3}}  \notag \\
G_{7_{\xi }}\left( \xi _{0}\right) &=&-\frac{3K}{\xi _{0}^{3}}+\frac{%
a_{1}\left\{ \mu _{\xi }\right\} _{0}}{\xi _{0}K}-\frac{a_{1}}{2\xi
_{0}^{2}K^{2}}-\frac{3a_{1}\left\{ \mu _{\xi }\right\} _{0}}{\xi _{0}} 
\notag \\
&&+\frac{a_{1}}{2\xi _{0}^{2}}+\frac{3}{\xi _{0}^{3}}  \label{ser23}
\end{eqnarray}

\section{$\qquad $Discussion and conclusion}

In the analysis above it has been shown how the problem of determining
slowly rotating sources embedded in Kerr empty space-time has reduced to the
problem of determining solutions of the system of ordinary differential
equations (\ref{eqn5}) to (\ref{eqn25}) subject to the conditions (\ref%
{ser16}) and (\ref{ser23}). Whilst these equations are under-determined they
may be supplemented by an equation of state to make unique solution
possible. Clearly the solution of the equations is straight forward enough
although analytic solution in closed form for any known $\lambda $ and $\mu $
seem not to be possible.

The most appropriate techniques for solution are either numerical or
alternatively a Taylor's series solution approach for $G_{i}\left( \xi
\right) $\ about $\xi =\xi _{0}$. Two variations on the series method are
worthy of note. In the first one may assume that both $\lambda $ and $\mu $
are known, for example, the Schwarzschild interior solution may be
specified, and then take the system (\ref{eqn5}) to (\ref{eqn25}) subject to
the conditions (\ref{ser16}) and (\ref{ser23}) as a linear set of equations
for $G_{i}\left( \xi \right) $. In the second one may assume that neither $%
\lambda $ and $\mu $ are known, then impose an appropriate equation of state
and also incorporate (\ref{stat10}) so as to develop a series solution for $%
\lambda $ and $\mu $.

It should be noted that solution of the equations (\ref{eqn5}) to (\ref%
{eqn25}) does not give the most general solution of the problem of
determining stationary rotating fluid sources which match Kerr empty
space-time. More general solutions may be obtained by appending terms of the
type:%
\begin{equation}
\sum_{k=2}^{\infty }Y_{k}\left( \xi \right) \sin ^{2k}\theta  \label{term4}
\end{equation}%
to each the substitutions (\ref{gees}) for $U$, $V$, $W$ and $H$ and then to
generate a further sequence of linear ordinary differential equations for $%
Y_{k}\left( \xi \right) $ similar to (\ref{eqn5}) to (\ref{eqn25}). In the
absence of an imposed equation of state then the resulting equations would
be solved subject to the conditions $Y_{k}\left( \xi _{0}\right) =0$ and
derivative $Y_{k_{\xi }}\left( \xi _{0}\right) =0$ for each value of $k$.
The equation of the boundary in case is (\ref{boun15}). When an equation of
state is imposed then the equation of the boundary is determined from (\ref%
{boun21}) and may contain higher harmonic component $\sin ^{2k}\theta $, $%
k\geq 2$ depending on the nature of the boundary expressions for $\left\{
U_{\xi }\right\} _{0}$ and $\left\{ V_{\xi \xi }\right\} _{0}$. These
expressions will in turn determine the boundary conditions which should be
imposed on $Y_{k}\left( \xi _{0}\right) $ and the derivatives $Y_{k_{\xi
}}\left( \xi _{0}\right) $.

The solution techniques outlined here are of course standard and are
therefore not pursued further. The relevance of the analysis to, for
example, determining potential sources for gravitational radiation or
determining more rapidly fluid sources matching empty space-time will be
examined in future research.

\end{document}